\documentclass[twocolumn,superscriptaddress,showpacs,preprintnumbers,amsmath,amssymb,doublespacing,prl]{revtex4}

\usepackage[dvips]{graphicx}
\usepackage{bm}

\newcommand{\msr}{$\mu$SR}
\newcommand{\cis}{CuIr$_2$S$_4$}
\newcommand{\czis}{Cu$_{1-x}$Zn$_x$Ir$_2$S$_4$}
\newcommand{\sro}{Sr$_2$IrO$_4$}

\begin{document} 

\title{Magnetic frustration in iridium spinel compound \cis}


\author{K.~M. Kojima}
\affiliation{Muon Science Laboratory and Condensed Matter Research Center, Institute of Materials Structure Science, High Energy Accelerator Research Organization (KEK), Tsukuba, Ibaraki 305-0801, Japan}
\affiliation{Department of Materials Structure Science, The Graduate University for Advanced Studies, Tsukuba, Ibaraki 305-0801, Japan}
\author{R. Kadono}\thanks{e-mail: ryosuke.kadono@kek.jp}
\affiliation{Muon Science Laboratory and Condensed Matter Research Center, Institute of Materials Structure Science, High Energy Accelerator Research Organization (KEK), Tsukuba, Ibaraki 305-0801, Japan}
\affiliation{Department of Materials Structure Science, The Graduate University for Advanced Studies, Tsukuba, Ibaraki 305-0801, Japan}
\author{M. Miyazaki}
\affiliation{Muon Science Laboratory and Condensed Matter Research Center, Institute of Materials Structure Science, High Energy Accelerator Research Organization (KEK), Tsukuba, Ibaraki 305-0801, Japan}
\author{M. Hiraishi}
\affiliation{Muon Science Laboratory and Condensed Matter Research Center, Institute of Materials Structure Science, High Energy Accelerator Research Organization (KEK), Tsukuba, Ibaraki 305-0801, Japan}
\author{I. Yamauchi}
\affiliation{Muon Science Laboratory and Condensed Matter Research Center, Institute of Materials Structure Science, High Energy Accelerator Research Organization (KEK), Tsukuba, Ibaraki 305-0801, Japan}
\author{\\A. Koda}
\affiliation{Muon Science Laboratory and Condensed Matter Research Center, Institute of Materials Structure Science, High Energy Accelerator Research Organization (KEK), Tsukuba, Ibaraki 305-0801, Japan}
\affiliation{Department of Materials Structure Science, The Graduate University for Advanced Studies, Tsukuba, Ibaraki 305-0801, Japan}

\author{Y. Tsuchiya}\thanks{Present address: Superconducting Wire Unit, National Institute for Materials Science, Tsukuba, Ibaraki 305-0003, Japan}
\affiliation{Quantum Beam Unit, National Institute for Materials Science,Tsukuba, Ibaraki 305-0003, Japan}
\author{H. S. Suzuki}
\affiliation{Quantum Beam Unit, National Institute for Materials Science,Tsukuba, Ibaraki 305-0003, Japan}
\author{H. Kitazawa}
\affiliation{Quantum Beam Unit, National Institute for Materials Science,Tsukuba, Ibaraki 305-0003, Japan}

\begin{abstract}
We demonstrate via a muon spin rotation experiment that the electronic ground state of the iridium spinel compound, \cis, is not the presumed spin-singlet state but a novel paramagnetic state, showing a quasistatic spin glass-like magnetism below $\sim$100 K. Considering the earlier indication that IrS$_6$ octahedra exhibit dimerization associated with the metal-to-insulator transition below 230 K, the present result suggests that a strong spin-orbit interaction may be playing an important role in determining the ground state that accompanies magnetic frustration.
\end{abstract}


\maketitle 

Geometrical frustration in electronic degrees of freedom such as spin, charge, and orbit, which is often realized in highly symmetric crystals, has been one of the major topics in the field of condensed matter physics.  Inorganic compounds with the AB$_2$X$_4$ spinel structure have offered fascinating insights from the viewpoint of their unusual physical properties relevant to geometrical frustration.  The thiospinel compound, \cis, is such a recent example, in which a charge order of mixed-valent Ir ions into isomorphic octamers of Ir$^{3+}_8$S$_{24}$ and Ir$^{4+}_8$S$_{24}$ with lattice dimerization of Ir$^{4+}$ pairs in the latter is realized upon metal-insulator (MI) transition at $T_{\rm MI}=$ 230 K\cite{Furubayashi:94,Matsuno:97,Nagata:98,Matsumoto:99,Burkov:00,Cao:01,Radaeli:02}. As per the currently accepted scenario regarding the $t_{2g}$ manifold, the frustration is relieved by the formation of Ir$^{4+}$ (5$d^5$, $S=1/2$) dimers that accompany the spin-singlet ground state driven by orbital order and the associated spin-Peierls instability\cite{Khomskii:05}. 

Meanwhile, it has been revealed in the iridium compound \sro\ that the crystal field levels of Ir$^{4+}$ ions in the insulating phase are reconstructed into a complex spin-orbital state represented by effective total angular momenta of $J_{\rm eff}=1/2$ and 3/2, where the half-filled $J_{\rm eff}=1/2$ level serves as a novel stage of the Mott transition due to on-site Coulomb interaction\cite{Kim:08,Kim:09}. The strong SO coupling entangles the spin and orbital degrees of freedom, where the magnetic interaction of corner-shared Ir$^{4+}$O$_6$ octahedra is modeled by the low-energy effective Hamiltonian consisting of terms representing the Heisenberg model, quantum compass model, and Dzyaloshinskii-Moriya (DM) interaction\cite{Jackeli:09}.  
The model has been successful in providing a microscopic account of the canted spin structure along the $xy$-plane in \sro\ as observed via resonant x-ray diffraction using $L$ edge ($2p\rightarrow 5d$)\cite{Kim:09}, wherein the interplay between the ``compass" and DM interactions is a crucial factor.

Interestingly, the same model predicts a completely different effective Hamiltonian for  {\it edge-shared}  Ir$^{4+}$O$_6$ octahedra, which is dominated by the quantum compass interaction\cite{Jackeli:09},
\begin{equation}
H^{(\gamma)}=-JS_i^\gamma S_j^\gamma,\label{compass}
\end{equation}
where $S_i^\gamma$ ($\gamma=x,y,z$) denotes the $\gamma$-component of the $S=1/2$ pseudospin operator $\vec{S}_i$ defined for the $J_{\rm eff}=1/2$ ground state, $J\simeq 2J_H/3U$ with $J_H$ and $U$ denoting the Hund's coupling and on-site Coulomb energy, respectively, and the entire energy is scaled by $4t^2/U$ with $t$ denoting the $dd$-transfer integral through an intermediate anion.  The form of exchange interaction depends on the spatial orientation of a given bond, thereby leading to a highly anisotropic (Ising type) interaction between Ir ions. 

\begin{figure}[h]
	\begin{center}
\includegraphics[width=0.95\linewidth]{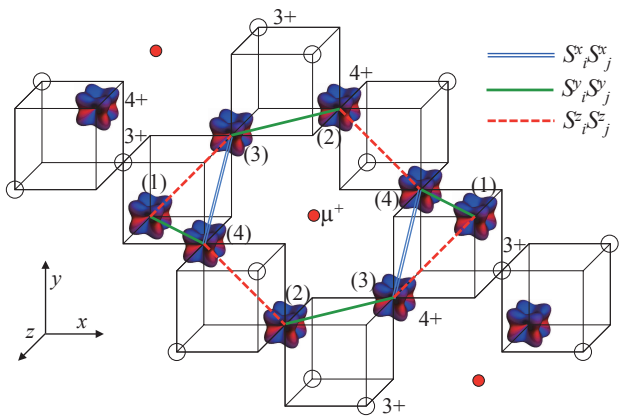}
			\caption{Octamer configuration associated with charge order in \cis. The exchange interaction $S_i^\gamma S_j^\gamma$ for each Ir$^{4+}$ pair is shown by lines along the respective $\gamma\gamma$ bond $(\gamma=x,y,z)$. The octupolar manifold at each Ir$^{4+}$ site represents the spin density profile in a hole of isospin-up state \cite{Jackeli:09}.  The bond length of the (1)-(4) and (2)-(3) Ir$^{4+}$ pairs is reported to shrink by $\sim$15\% upon charge ordering and the associated structural transition \cite{Radaeli:02}.}
			\label{fig1}
	\end{center}
\end{figure}

It is noteworthy that upon substituting IrO$_6$ with IrS$_6$, \cis\ in the insulating phase may serve as an ideal stage of the quantum compass model, as it partly comprises an edge-shared Ir$^{4+}$S$_6$ network. More importantly, the compass interaction is projected to the Kitaev model on the Ir$^{4+}$ octamers\cite{Jackeli:09,Kitaev:06} in the case of charge order without dimerization.  As shown in Fig.~\ref{fig1}, an Ir$^{4+}$ octamer consists of a honeycomb unit and two triangular units, on which every Ir$^{4+}$ ion occupies a node of two (or three) nonequivalent bonds with each being perpendicular to one of the cubic axes $x$, $y$, $z$.  Thus it is speculated that the iridium pseudospins (accompanying effective moments) are subject to strong geometrical frustration due to competing exchange interactions favoring different spatial orientations.

From this revised viewpoint, one alternative scenario to spin-singlet formation is to understand the lattice dimerization and associated structural phase transition in \cis\ as a relief of magnetic frustration upon charge ordering. In this regard, it would be interesting to note that dimerization occurs for Ir$^{4+}$ pairs connected by one particular kind of bond on an octamer\cite{Radaeli:02}, e.g., $yy$ or $zz$ bond (shared by triangular units) in Fig.~\ref{fig1}. This suggests that, while the Kitaev ground state might be destroyed by the lattice dimerization (due to enhanced exchange interaction for the corresponding bond), the magnetic frustration is only partially removed.  In any case, such a situation would be drastically different from the spin-singlet ground state predicted within the conventional scenario where spins are independent of orbital degrees of freedom, and it may call for reconsideration of the origin of MI transition.

Here, we report muon spin rotation (\msr) measurements of powder samples of \cis\ and related compounds that establish the development of a highly disordered weak magnetism at low temperatures.  Evidence of residual paramagnetism below $T_{\rm MI}$ is also provided by an anomalous chemical shift in $^{63}$Cu-nuclear magnetic resonance (NMR). A detailed analysis of \msr\ data suggests that the magnetism is associated with Ir$^{4+}$ ions located in the triangular units of the octamers. These results imply that lattice dimerization does not accompany ``spin dimerization" to form the singlet-ground state, suggesting the importance of SO coupling
in understanding the unusual local magnetism. 

Typical examples of \msr\ spectra [time-dependent $\mu$-e decay asymmetry, $A(t)$] in \cis\ at various temperatures are shown in Fig.~\ref{fig2}a (details regarding the samples and \msr\ experiment are found in Supplementary Information). We can observe a slow Gaussian damping at 200 K, which is expected for muons exposed to random local magnetic fields from nuclear magnetic moments (primarily from $^{63}$Cu and $^{65}$Cu in \cis).  Below $\sim$100 K, fast exponential depolarization sets in, where the depolarization rate ($\lambda$) as well as the relative amplitude ($a$) of the depolarizing component increases with decreasing temperature.  These spectra are analyzed by a curve fit from an equation of the form
\begin{equation}
A(t)\simeq A_0[a e^{-(\lambda t)^\beta}+ (1-a)]G_z^{\rm KT}(\Delta t),\label{asy}
\end{equation}
where $A_0$ denotes the initial asymmetry, $\beta$ denotes the power for the ``stretched" exponential damping, and $G_z^{\rm KT}(\Delta t)$ ($\simeq e^{-\Delta^2t^2/2}$ for $\Delta t\ll 1$) denotes the Kubo-Toyabe relaxation function with $\Delta$ representing the line width determined by the nuclear dipolar fields\cite{Hayano:79}. Here, we adopted the stretched exponential damping as a working model to describe the behavior of \msr\ time spectra associated with highly disordered magnetism\cite{Mekata:00}, which turns out to provide an excellent curve fitting to the measured data with a tendency of smaller $\beta$ at lower temperatures. 
A comparison of the deduced value of $\Delta$ [$=0.150(1)$ $\mu$s$^{-1}$] with the calculated second moments of Cu and Ir nuclei leads us to conclude that the muon site is located at the center of the Ir$^{3+/4+}$ octamers [(1/8,1/8,1/8) $O_h$ site situated at the center of a S$^{2-}_6$ octahedron, as shown in Fig.~\ref{fig1}, where $\Delta$ is calculated to be 0.152 $\mu$s$^{-1}$].  (As shown in the Supplementary Fig.~1, the site corresponds to the saddle point of the calculated $\Delta$, which assures the uniqueness of the site assignment.) The nearly exponential damping in the time domain corresponds to a Lorentzian-type distribution of the internal magnetic field at the muon site, thereby indicating the onset of highly disordered magnetism. The response of the \msr\ spectra to a longitudinal field (parallel to initial muon polarization, Fig.~\ref{fig2}b) implies that the internal field is quasistatic at 2 K within the time window of \msr\ ($<10^{-5}$ s).
The form within the square brackets in Eq.~(\ref{asy}) is subsequently replaced with the physically more meaningful form obtained as an approximation of the relaxation function for spin glass\cite{Kubo:81}, 
\begin{equation}
\simeq f\left\{\frac{1}{3}+\frac{2}{3}e^{-(\lambda t)^\beta}\right\}+(1-f),\label{sed}
\end{equation}
where $f$ [$=0.435(5)\simeq3a/2$ at 2 K] represents the true fractional yield for muons probing the inhomogeneous magnetic field $B$ whose density distribution is represented by, 
\begin{equation}
P(B)\sim \int_{-\infty}^{\infty}e^{-(\lambda t)^\beta}\cos(\gamma_\mu Bt)\:dt,\label{fr}
\end{equation}
with $\gamma_\mu$ ($=2\pi\times13.553$ MHz) denoting the muon gyromagnetic ratio. 
While Eq.~(\ref{fr}) corresponds to the Lorentzian distribution for $\beta=1$, the observed tendency of $\beta<1$ (see inset of Fig.~\ref{fig2}d) suggests that $P(B)$ has a broad distribution that we shall briefly refer to as super-Lorentzian. This, together with the absence of oscillatory behavior\cite{Kubo:81} in Eq.~(\ref{sed}) suggests notable difference of the ground state from that of the conventional spin glass, calling for more detailed investigation in the future.

\begin{figure}
	\begin{center}
\includegraphics[width=0.95\linewidth]{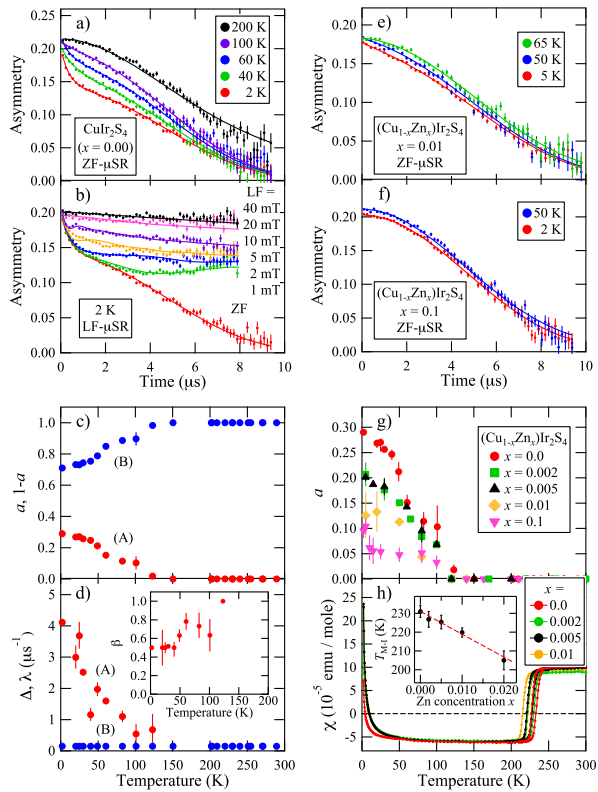}
			\caption{(a) Time-dependent \msr\ spectra observed at several temperatures in powder sample of \cis\ ($x=0$) under zero external field, with solid curves showing fits by Eq.~(\ref{asy}). (b) \msr\ spectra at 2 K under longitudinal fields of 1, 2, 5, 10, 20, and 40 mT, where the solid curves are fits using the Lorentzian Kubo-Toyabe relaxation function for the $f$ component.  (c) Temperature dependence of the partial $\mu$-e decay asymmetry, one for the component showing fast exponential damping (A) and the other showing Gaussian damping (B). (d) Depolarization rate of the respective components [with (A) and (B) representing the same in (c)] as a function of temperature.
 (e) \msr\ spectra in \czis\ under zero external field with $x=0.01$, and (f) $x=0.1$, where solid curves show fits by Eq.~(\ref{asy}). (g) Relative amplitude of the component showing fast exponential damping (A) as a function of temperature. (h) Magnetic susceptibility ($\chi$) of \czis\ samples used for \msr\ measurements (measured at 0.1 T). Inset shows the temperature of metal-insulator transition ($T_{\rm MI}$) determined by $\chi$. [Vertical axes for (a), (b), (e), and (f) are scaled by $A_0$ in Eq.~(\ref{asy}) for the ease of comparison by eye.] }
			\label{fig2}
	\end{center}
\end{figure} 

Since muons surrounded only by non-magnetic Ir ions at the six nearest-neighboring (nn) sites [comprising Ir(2)-(3)-(4) cyclic hexamers, see Fig.~1] would exhibit slow Gaussian damping due to nuclear random local fields (corresponding to the component $1-f$), the value of $f$ imposes a constraint on the spin  configuration of the hexamer.  It should be noted that there are 16 muon sites in the unit cell of the charge-ordered state with (Ir$^{3+})_n$(Ir$^{4+})_{6-n}$ hexamers being classified by $n=0$ (1 site), 2 (7 sites), 4 (7 sites), and 6 (1 site) in terms of the Ir charge state, where the site with $n=6$ is the only one that consists of non-magnetic nn Ir$^{3+}$ ions.  Provided that a muon occupies these sites statistically at random so that the probability for $n=6$ would be 1/16, the magnitude of $f$ actually implies that a certain fraction of Ir$^{4+}$ ions besides Ir$^{3+}$ ions must be non-magnetic. More detailed crystallographic analysis suggests that Ir$^{4+}$(3) or Ir$^{4+}$(4) ions bear a local magnetic moment on the hexamer, and correspondingly, 6 out of 16 muon sites would become magnetic (i.e., $f\simeq6/16$): this hypothesis is in reasonable agreement with the observed magnitude of $f$.

The highly inhomogeneous magnetic ground state of Ir$^{4+}$ octamers suggests the presence of frustration and associated degeneracy in the electronic ground state of \cis, which is in favor of the ``revised view" where the SO interaction might lead to geometrical frustration.   A crude estimation can be made for the Ir$^{4+}$ effective moment size as a mean value by comparing $\lambda^\beta$ [$=2.03(8)$ MHz at 2 K] with the hyperfine parameter ($\delta_\mu$) calculated using the second moment of the magnetic fields from $\vec{S_i}$. Using a dipolar field approximation, we obtain $\delta_\mu=24.1$ $\mu$s$^{-1}/\mu_B$ for two Ir$^{4+}$ moments, which yields $\overline{m}=\lambda^\beta/\delta_\mu=0.085(3)\mu_B$.  This small moment size is in line with the extremely small magnetic susceptibility below $T_{\rm MI}$ that makes it difficult to detect signals associated with local spin magnetism by the conventional ac-$\chi$ measurement (see Fig.~\ref{fig2}h, where $\chi$ seems to be predominantly determined by a diamagnetic offset $\chi_{\rm dia}\simeq-6\times10^{-6}$ emu/mol with little variation of $|\chi-\chi_{\rm dia}|\le10^{-6}$ emu/mol over 50--200 K).

The presence of residual paramagnetism is also evidenced by anomalous chemical shift ($^{63}K$) of $^{63}$Cu-NMR associated with the MI transition (see Supplementary Information).  We infer from the comparison between $^{63}K$ and uniform susceptibility ($\chi$) in the metallic phase of \cis\ that the hyperfine coupling of $^{63}$Cu nuclei is $A_{\rm hf}=7\pm2$ T/$\mu_B$ with a positive sign. Although the magnitude of $A_{\rm hf}$ would have considerable uncertainty as it was deduced from the small shifts of  $^{63}K$ and $\chi$, there should be least ambiguity in its sign.  This leads us to expect that $^{63}K$ would exhibit a negative shift upon formation of the spin-singlet ground state in the insulating phase. In contrast, the observed positive shift of $^{63}K$ below $T_{\rm MI}$ (Supplementary Fig.~2b) thus indicates that a certain paramagnetism develops below $T_{\rm MI}$.  Considering that the {\sl positive} chemical shift can be primarily attributed to the orbital component of 5$d$ electrons, this result suggests that the anomaly originates from the presence of an unconventional paramagnetic ground state under strong SO interaction.  It is readily understood that such a paramagnetic state serves as a stage of observed weak random magnetism probed by \msr.

It should be noted that the occurrence of weak magnetism has been already hinted in some of the earlier literatures.  
For example, the spin-lattice relaxation rate ($1/T_1$) versus temperature ($T$) plot in an earlier report of $^{63}$Cu-NMR studies exhibits an unusual power law, $(T_1T)^{-1}\propto T^2$ in the insulating phase where thermal activation behavior ($\propto e^{-W/k_BT}$, with $W$ being the band gap) would be anticipated\cite{Kumagai:04}. While this peculiar result is currently attributed to an anisotropic band structure, it may suggest the absence of the excitation gap expected for the spin-Peierls state. We also point out that the additional broadening of the quasistatic linewidth by $\lambda^\beta/\gamma_\mu\sim$0.003 mT below $\sim$100 K may not have been clearly identified in the powder spectra of $^{63}$Cu-NMR due to the complicated line shape originating from quadrupole splittings.

The magnetic ground state exhibits another intriguing feature in that the quasistatic local magnetism is strongly suppressed by the substitution of Cu by Zn.  As shown in Figs.~\ref{fig2}e-g, Zn content as small as $x=0.01$ is sufficient to eliminate the signal showing exponential damping. Meanwhile, the magnetism appears to be robust to changes in the off-stoichiometric composition of Ir and Cu. We confirmed that \msr\ measurements of CuIr$_{1.96}$S$_4$ and Cu$_{0.98}$Ir$_2$S$_4$ samples perfectly reproduced the result of pristine \cis\ (see Supplementary Fig.~3).

The Zn substitution of Cu in \czis\ is known to drastically reduce the charge order and associated structural phase transition from cubic to triclinic with decreasing temperature (see Fig.~\ref{fig2}h), eventually leading to the formation of a metallic phase that exhibits superconductivity for samples with $0.25\le x\le0.8$ \cite{Cao:01}. The change in the Pauli paramagnetic susceptibility indicates the presence of a hole-filling mechanism due to an excess electron from Zn (Cu$^{1+}\rightarrow$ Zn$^{2+}+e^-$). Assuming that $x\simeq0.01$ is the critical Zn content, one hole is sufficient to influence the electronic correlation of 12 Ir octamers.  This unusual sensitivity of the ground state to hole-filling supports the idea that there exists an inter-octamer correlation.  
It is also inferred from a numerical study of the Heisenberg-Kitaev model \cite{Chaloupka:10} that an isolated Ir$^{4+}$ octamer may not retain magnetic moments regardless of partial dimerization, thereby suggesting that the other interactions not included in the model form a crucial factor in stabilizing the spin-glass-like ground state \cite{Nasu:13}.
The presence of local magnetic moments on Ir$^{4+}$(3) or Ir$^{4+}$(4) may suggest that the inter-octamer correlation can be mediated through the triangular units of the octamers. 
The inter-octamer correlation would also provide a qualitative explanation for the superstructure (i.e., alternative stacking) of Ir$^{3+}$ and Ir$^{4+}$ octamers observed in \cis\cite{Radaeli:02}.  

Finally, we discuss the possible extrinsic effects associated with muon implantation.  It has been reported that x-ray irradiation of \cis\ induces a modulation of its crystal structure from triclinic to tetragonal below $\sim$50 K\cite{Ishibashi:02}.  This gives rise to the concern that muon irradiation might have caused similar modulation to which the emergence of disordered magnetism may be attributed. However, it is inferred from the diffuse scattering analysis of the x-ray modulated phase that the local structure in the tetragonal phase still remains triclinic, thereby strongly suggesting that the Ir$^{4+}$ dimerization and associated octamer structure is preserved\cite{Ishibashi:02}.  Thus even if the muon irradiation has a similar influence on the crystal, we can reasonably assume that the local electronic correlation of Ir$^{4+}$ octamers remains virtually intact. We also note that the escape length of implanted muons from the end of the radiation track in matter (where the energy loss process transforms from local electronic excitation to elastic scattering with host atoms) is estimated to be 10$^{1}$ $\mu$m in order of magnitude\cite{Zeigler:85}, so that muon would not be affected by the radiolysis product.

Another concern is the effect of the positive charge introduced by the muon.  It is known that the muon behaves as a light interstitial proton or hydrogen (p$^+$ or H) in matter, where the local electronic structure associated with the muon is equivalent to that of p$^+$/H (except for a small correction of $\simeq$0.5\% due to the difference in the reduced mass).  It is empirically well established that implanted muons  tend to reside at interstitial sites close to anions where the electrostatic potential is minimal, forming an O-H bond in oxides.   In the present case of \cis, the muon is situated at the center of the S$^{2-}_6$ octahedron, where the positive charge is effectively screened by anions.  Moreover, the high symmetry of the muon site makes it improbable that the muon exerts an asymmetric electric field gradient on the surrounding Ir ions in the octamer. Thus, even when there is some energy offset of electronic levels induced by the muon, it would be common to all the Ir$^{4+}$ ions in an octamer, and therefore, the primary hypothesis underlying the physics based on the strong SO coupling can still be considered valid.

 We would like to thank the staff of TRIUMF and J-PARC MUSE for their technical support during the $\mu$SR experiment. Thanks are also due to A. Nakao, H. Nakao, and R. Kumai for performing a detailed x-ray diffraction study on our \czis\ sample.  We appreciate the stimulating discussions we had with K. Kumagai, M. Yoshida, M. Takigawa, T. Arima, and Y. Motome.

\end{document}